\documentclass{article}
\usepackage[utf8]{inputenc}
\usepackage{authblk}
\usepackage{graphicx}
\graphicspath{{noiseimages/}}
\usepackage{xcolor}
\usepackage[T2A]{fontenc}
\usepackage{amssymb}
\usepackage{tikz}
\usepackage{pgfplots}
\usepackage{mathrsfs}
\pgfplotsset{compat=1.7}
\usetikzlibrary{intersections, pgfplots.fillbetween}
\usetikzlibrary{decorations.pathmorphing}
\usetikzlibrary{arrows.meta}
\usepackage[mode=buildnew]{standalone}
\usepackage[toc,page]{appendix}
\usepackage{amsmath}
\usepackage{bm}
\usepackage{float}
\usepackage{comment}
\usepackage{a4wide}
\usepackage[english]{babel}
\usepackage{hyperref}

\definecolor{urlcolor}{HTML}{990000}
\definecolor{linkcolor}{HTML}{005F5F}
\hypersetup{
    pdfstartview=FitH,
    linkcolor=linkcolor,
    urlcolor=urlcolor,
    colorlinks=true,
    citecolor=blue
}

\author[1,2]{D. V. Diakonov\footnote{\tt dmitrii.dyakonov@phystech.edu}}

\author[1]{Gleb S. Zverev\footnote{\tt zverev.gs@phystech.edu}}

\affil[1]{Institutskii per. 9, Moscow Institute of Physics and Technology,
141700, Dolgoprudny, Russia}
\affil[2]{\itshape Bol'shoi Karetnyi per., 19, Institute for Information
Transmission Problems, 127994, Moscow, Russia}

\title{\textcolor{black}{Entanglement entropy of black holes from the first law of thermodynamics}}
\date{}

\begin{document}
\maketitle

\begin{abstract}
We compute the entanglement entropy of quantum matter in spacetimes with a Killing horizon entirely on-shell, by integrating the first law of thermodynamics along a one-parameter family of equilibrium backgrounds labelled by the Hawking temperature. This yields a closed expression in terms of the renormalized stress-energy tensor and the Killing vector, finite by construction, with no replicas and no conical singularities. Its covariant form agrees with the matter terms of the Bardeen--Carter--Hawking mass variation formula. We show in details that for  static patch of de Sitter space and for the non-rotating BTZ black hole it reproduces the conical entropy exactly, contact term $Q$ included.
\end{abstract}

\newpage

\section{Introduction}

The laws of black hole thermodynamics, formulated in the works \cite{Bekenstein:1972tm, Bekenstein:1973ur, Bardeen:1973gs}, are understood to have a standard thermodynamic interpretation. In particular, the first law for an uncharged black hole represents an analogue of the thermodynamic first law:
\begin{equation}\label{BH_thermo}
    \delta M = \frac{\kappa}{8\pi}\delta A + \Omega_{H}\delta J,
\end{equation}
where $M$ and $J$ are the black hole parameters (mass and angular momentum, respectively), $A$ is the horizon area, and $\kappa$ is the surface gravity. Following the classic results of \cite{Bekenstein:1972tm, Hawking:1975vcx}, the first term in (\ref{BH_thermo}) is interpreted in terms of the Bekenstein-Hawking temperature $T_{H} = \frac{\kappa}{2\pi}$ and the black hole entropy $S_{BH} = \frac{A}{4}$, which are associated with the surface gravity and the horizon area.

If a black hole is surrounded by classical matter, then the total entropy is given by the generalized entropy:
\begin{equation}\label{gen_td_entropy}
    S_{tot} = \frac{\bar{A}}{4G} + S_{matt},
\end{equation}
where $\bar{A}$ is the horizon area of a black hole which takes into account the back-reaction and $S_{matt}$ represents the entropy of the quantum matter outside the horizon. It is worth noting that equation (\ref{gen_td_entropy}) represents a special case of a more general statement \cite{Lewkowycz:2013nqa} about generalized gravitational entropy. Specifically, the first term within equation (\ref{gen_td_entropy}) corresponds to the area of a minimal codimension two hypersurface, which, in the context of a black hole, corresponds to the horizon.

In black hole thermodynamics, the entanglement entropy is typically computed using off-shell methods such as the replica trick \cite{Callan:1994py, Holzhey:1994we}, the brick wall model \cite{tHooft:1984kcu}, and others (see reviews \cite{Solodukhin:2011gn, Witten:2024upt}). The term "off-shell" here means that one considers the case where the temperature $\beta^{-1}$ of the system is not equal to the black hole's Hawking temperature $\beta^{-1}_{H}$. This immediately leads to a problem because no regular, complete Euclidean solution is available. To work around this, one must either use a background metric that does not satisfy the vacuum gravitational equations, or exclude some region near the horizon, thereby making the solution incomplete. In both cases, the calculation of the free energy encounters difficulties, and the result may depend on the particular off-shell procedure chosen. This naturally leads to the question:
\begin{center}
    \textbf{Why are off-shell methods needed at all to compute black hole entanglement entropy?}
\end{center}

The main purpose of our paper is to establish the relation between off-shell calculations of black hole entanglement entropy and the thermodynamic black hole entropy that arises from the on-shell first law of thermodynamics. "On-shell" means that all observables characterizing a black hole are in thermal equilibrium. We compare a one-parameter family of classical equilibrium backgrounds labelled by their Hawking temperature, and then integrate the first law along it. No off-shell geometry is involved, and no backreaction is needed at one loop approximation at this stage, since the backreaction of quantum matter enters separately, as a shift of the horizon area in the gravitational part of the generalized entropy. As a result, the entanglement entropy can indeed be computed on-shell:
\begin{equation}\label{Our_Formula}
 \boxed{S^{on-shell}_{matt}=\int^{T_H} \frac{d T_H}{T_H} \int
  \left(\frac{1}{2}\,\langle T^{\mu\nu}\rangle\,\frac{\partial g_{\mu\nu}}{\partial T_H}\,
  \xi^\rho\, d V_\rho
  -\frac{\partial}{\partial T_H}\Big(\langle T^{\mu}_{\ \nu}\rangle\,\xi^\nu\, dV_\mu\Big)
  \right)},
\end{equation}
where we use the renormalized expectation value of the stress-energy tensor taken over the thermal state at the Gibbons–Hawking temperature. Crucially, the background metric itself is a function of this temperature $ g_{\mu \nu} (T_H).$

Note that this expression is in agreement with the generalized ADM mass variation formula for a stationary axisymmetric black hole surrounded by classical matter \cite{Bardeen:1973gs}:
\begin{align}
    \delta \bar{M} =\frac{\kappa}{8\pi}\delta \bar{A} + \Omega_{H}\delta J +\frac{1}{2} \int T^{\mu\nu} \delta g_{\mu\nu} \xi^\rho d V_\rho-\delta \int T^\mu_\nu \xi^\nu dV_\mu,
\end{align}
where quantities with a bar represent corrections due to the matter field. In this equation, the last two terms can be interpreted as the variation of the entanglement entropy, i.e., $T_H \delta S^{ent}$. 
Note that the variation is taken inside the integral. But we use a form in which the matter terms arise in the covariant phase space derivation of the first law \cite{Iyer:1996ky,Banihashemi:2022htw}, where $\Sigma$ is a fixed submanifold and $\delta$ acts on the fields and on the volume form. Hence we can write $\int \partial_{T_H}$ instead $\partial_{T_H}\int$, as in the ADM mass variation formula, requires a choice of the diffeomorphism identifying the two solutions by keeps the horizon at a fixed coordinate location, i.e. $\bar r = r/r_h$ with the horizon at $\bar r = 1$, we will discuss that point in Section~\ref{sec:onshell}.

Before moving on to the derivation of formula (\ref{Our_Formula}), its verification with two examples, and its comparison with off-shell entropy calculations, we will first discuss the general differences between the statistical-mechanical and thermodynamic approaches.

For a black hole, one can define several distinct notions of entropy. First, there is the thermodynamical entropy, an on-shell quantity derived from the partition function:
\begin{equation}
    S^{TD} = -\left(\beta_H \frac{\partial}{\partial \beta_{H}}-1\right)\log{Z(\beta_{H})}.
\end{equation}
Second, there is the statistical mechanics (von Neumann) entropy, which is based on the density matrix of the system:
\begin{equation}
    S^{SM}= -\text{Tr}{\rho \log \rho}.
\end{equation}
Finally, there is the conical singularity entropy, an off-shell quantity:
\begin{equation}
    S^{con} = -\left(\beta \frac{\partial}{\partial \beta}-1\right) \log{Z({\beta})}\Big|_{\beta=\beta_H} = - \frac{d}{d\alpha}\text{Tr}\rho^{\alpha}\bigg|_{\alpha=1}.
\end{equation}
 In what follows, we will distinguish between the contributions arising semiclassically—namely, the contributions from the gravitational action $(gr)$ and the contribution from matter $(matt)$—and the total contribution, denoted as $(tot)$.
The Bekenstein-Hawking entropy $S^{BH}_{gr}=A/4$ originally appears within the thermodynamical framework \cite{Bekenstein:1972tm, Bekenstein:1973ur, Bardeen:1973gs} as $S^{BH}_{gr}=S^{TD}_{gr}$. One of the main problems in black hole physics is to find its statistical-mechanical foundation; that is, to define the internal degrees of freedom, find the corresponding density matrix, and establish the relation between the thermodynamic and statistical-mechanical definitions such that $S^{TD}_{gr}=S^{SM}$. While there has been progress in various models, this remains an open problem. The general assumption is that, as a compact object, a black hole possesses microstates hidden behind the horizon whose degeneracy can be counted; for more details, see the review \cite{Carlip:2014pma}. A particularly significant development was the work \cite{Strominger:1996sh}, which showed that the entropy of extreme black holes in string theory, calculated from the counting of BPS states, exactly matches the Bekenstein-Hawking formula for large charges. Another equally significant "central dogma" is expressed in the paper \cite{Almheiri:2020cfm}, in which the authors propose that, from the outside, a black hole can be understood as a quantum system with $\frac{A}{4G}$ degrees of freedom that evolves unitarily under time evolution.

It is important to note that for the semiclassical case the conical method gives exactly the thermodynamical entropy, $S^{con}_{gr}=S^{TD}_{gr}$ \cite{Solodukhin:1996vx}. Moreover, as shown in \cite{Wald:1993nt, Jacobson:1993vj, Iyer:1994ys, Jacobson:1994qe}, the Bekenstein-Hawking entropy can be derived as an integral of the Wald Noether charge, a result that holds in any generalized theory of gravity: $S^{Wald}_{gr}=S^{BH}_{gr}$. Furthermore, the conical singularity method yields precisely this Wald entropy: $S^{con}_{gr}=S^{Wald}_{gr}$ \cite{Solodukhin:2011gn}.

When it comes to quantum corrections to black hole entropy, the situation differs significantly from the semiclassical case. A variety of off-shell methods exist for computing these corrections, and they often yield different results (for a detailed review, see \cite{Frolov:1998vs,Solodukhin:2011gn}). In this discussion, we focus specifically on the conical singularity method. Unlike some other approaches that impose a cutoff near the horizon—thereby omitting essential physics and providing an incomplete description—the conical singularity method preserves the near-horizon region. To properly describe quantum corrections, one should begin with the pure vacuum state and trace out the degrees of freedom of the quantum field inside the horizon. This procedure yields a thermal density matrix $\hat{\rho} = e^{-\beta \hat{H}}$ for the exterior, from which we can compute the statistical mechanical entropy $S^{SM} _{matt}= -\operatorname{Tr} \hat{\rho} \log \hat{\rho}$. By construction, this quantity is precisely the entanglement entropy associated with the horizon $\Sigma$. Moreover, because the resulting density matrix is thermal, this entanglement entropy coincides with the entropy of the black hole's thermal atmosphere. It is worth noting that numerous attempts have been made to connect the Bekenstein-Hawking entropy to this thermal atmosphere. In standard thermodynamics, the thermodynamic and statistical mechanical entropies are exactly equivalent: $S^{TD} = S^{SM}$. However, as argued in \cite{Frolov:1994zi}, black holes may represent a notable exception to this rule, serving as an example where these two entropies do not coincide \cite{Frolov:1995xe}:
\begin{align}
    S^{TD}_{matt} = S^{SM}_{matt}+ \Delta S,
\end{align}
where $\Delta S$ is an additional finite correction caused by the shift of the black hole horizon because of quantum effects.

As a result, for a minimally coupled theory with a vanishing cosmological constant, the relation between the thermodynamical black hole entropy and the statistical mechanical entropy is given by \cite{Fursaev:2022goi}
\begin{align}
    S^{TD}_{tot}=\bar{S}^{BH}_{gr}+S^{SM}_{matt}.
\end{align}
Here, $S^{TD}_{tot}$ contains both classical (tree-level) and quantum contributions, while $\bar{S}^{BH}_{gr}$ is the Bekenstein-Hawking entropy with a shifted horizon area due to quantum backreaction. That also means that if we take into account backreaction then there is a quantum correction to the Bekenstein-Hawking temperature:
\begin{align}
    \bar{T}_{H}= T_{H}+\delta T_{H},
\end{align}
and then we should take this into account when we compute $S^{con}_{tot}$. At one loop, it must be taken into account only for the gravitational part $S^{con}_{gr}$, since the matter contribution $S^{con}_{matt}$ is already the leading correction, while taking into account corrections to the temperature would give subleading corrections. As a result, the conical method gives:
\begin{align}
    S^{TD}_{tot}= S^{con}_{gr}\Big|_{\beta=\beta_H+\delta \beta_H}+S^{con}_{matter} \Big|_{\beta=\beta_H}
\end{align}

For a minimally coupled theory, there are formal arguments that the statistical mechanical entropy coincides with the entropy appearing in the conical approach, i.e., $S^{con}_{matt}=S^{SM}_{matt}$ \cite{Solodukhin:1996vx}. However, for a non-minimally coupled theory, such as $\mathcal{L}=...+1/2\xi R \phi^2$, a contact term $Q$ appears in the relation:
\begin{align}
    S^{con}_{matt}=S^{SM}_{matt}-Q,
\end{align}
where:
\begin{align}
   Q= \beta_H \xi \int_{\Sigma} \langle \phi^{2} \rangle .
\end{align}
This term arises from the conical singularity contribution to the curvature \cite{Solodukhin:1995ak}:
\begin{align}
    R_\alpha= R + 4\pi(1-\alpha)\delta(\Sigma)
\end{align}
and is necessary for a self-consistent renormalization procedure of the effective action and entropy. This same contact term also appears in the difference between the energy $E=-\langle T^0_0\rangle$, which contributes to the total mass of the black hole, and the canonical energy $H$, which generates evolution along the Killing time \cite{Frolov:1998vs}:
\begin{align}
    \langle \hat E\rangle-\langle \hat H\rangle=-T_H\, Q.\label{EHQ}
\end{align}
A major issue is that this contact term $Q$ lacks a straightforward statistical explanation for scalar field theory \cite{Solodukhin:2011gn,Fursaev:2022goi}. It can, however, be reinterpreted  1) as a contribution to the Wald entropy \cite{Donnelly:2012st}, i.e., $Q$ is a Noether charge analogous to Wald's charge \cite{Wald:1993nt}, 2) as a contribution of soft modes that are located in the nearest vicinity of the horizon \cite{Frolov:1997up}. Similar contact terms arise naturally in gauge theories, where they have been interpreted as the statistical-mechanical entropy of edge modes living on the boundary \cite{Donnelly:2012st,Blommaert:2018rsf,Ball:2024hqe}, that arise from residual gauge symmetries.

As a result, the total thermodynamical black hole entropy can be expressed as a sum of three terms \cite{Fursaev:2022goi}:
\begin{equation}
    S^{TD}_{tot}=\bar{S}^{BH}_{gr}+S^{SM}_{matt}-Q.
\end{equation}
This relation was illustrated in a two-dimensional model, where all the calculations can be performed exactly \cite{Frolov:1995xe} with the restriction that the cosmological constant vanishes. In section \ref{sec:dS} we show that in de Sitter this relation holds, but in the other sense, i.e. with the definition that $S^{TD}_{tot}=\log Z_{total}$.

It is well known that entanglement entropy in conical approach $S_{matt}^{con}$ is divergent, with the leading term proportional to the area of the horizon surface \cite{Kabat:1995eq}, $S_{matt}^{con}\sim A/\epsilon^2$. This divergence can be absorbed by a renormalization of Newton's constant in $S^{BH}_{gr}$. Sub-leading divergences can similarly be renormalized by redefining the higher-curvature couplings that contribute to the black hole entropy. In addition, one must also remove the standard ultra-violet divergences in the effective action. Both of these renormalization procedures are self-consistent for minimally coupled field theories \cite{Solodukhin:1995ak}. For non-minimally coupled theories, one must also take into account the term $Q$, which itself has a divergent contribution; including it, the renormalization procedure is self-consistent. It is also important to note that the conical method is not suitable for computing second derivatives of the free energy, such as those needed for the heat capacity, because at second order the free energy contains a contribution that behaves like $\int dV\delta^2(\Sigma)$ \cite{Solodukhin:2011gn}. Consequently, it is unclear whether the standard renormalization procedure remains well-defined at higher orders. Further problems arise when self-interactions are switched on; as argued in \cite{Solodukhin:2011gn}, the renormalization procedures for the entanglement entropy and the effective action are then no longer self-consistent.

The local thermodynamic approach demonstrates an equivalence between the on-shell calculation and the conical method, which yields a sum of the statistical mechanical contribution and the contact terms:
\begin{align}
    S^{on-shell}_{matt}=S^{con}_{matt}=S^{SM}_{matt}-Q.
\end{align} 
We therefore provide a method for computing the entanglement entropy using an on-shell approach. Moreover, since our method uses the renormalized value of the stress-energy tensor, one immediately obtains a finite, renormalized value for the entanglement entropy; hence one can also use this method to calculate the renormalized value of any thermodynamic quantity. Loop corrections to the entropy due to self-interactions can also be obtained straightforwardly, as the renormalization of the stress-energy tensor is well defined. Hence the local thermodynamics approach solves both problems.

Finally, we address a common naive argument that entanglement entropy depends only on the area of the entangling surface and its geometric characteristics, and is therefore not an extensive quantity. Indeed, by dividing a system into a subsystem $B$ and its complement $\bar{B}$, one can show that $S(B) = S(\bar{B})$, regardless of how small $B$ is. From this, one might assume that the entropy must depend on something common to both $B$ and $\bar{B}$, the only common element being their shared boundary $\partial B = \partial\bar{B}$. However, from the perspective of our on-shell method, the entanglement entropy $S^{ent}=S^{on-shell}$ is expressed as a volume integral, $S^{on-shell}\sim \int dV$. It is worth noting that even in the off-shell approach, one can show that the statistical mechanical entropy in a space-time with a Killing horizon can be interpreted as arising from either bulk or boundary contributions, $S^{SM}\sim \int dV \partial (\cdots) =\int dA (\cdots)$, precisely because the integrand is a total derivative \cite{Diakonov:2023jdk}.

\newpage
\section{Off-shell approach: Replica trick}\label{sec:offshell}
We begin with a brief review of the off-shell technique for computing the entanglement entropy based on the replica trick (for more details see \cite{Callan:1994py, Holzhey:1994we, Calabrese:2004eu} and also the reviews \cite{Solodukhin:2011gn,Nishioka:2018khk}).

\subsection{Thermofield double and replicas}

Consider a QFT on a Euclidean manifold $\mathcal{M}$ with periodic Euclidean time $\tau \to \tau +\beta_{H}$. Let the Hilbert space of the system factorize as $\mathcal{H} = \mathcal{H}_{A}\otimes\mathcal{H}_{B}$, and let the wave functional of the theory be the entangled thermofield double state
\begin{equation}
    \Psi_{TFD}[\phi_{A},\phi_{B}] = \frac{1}{\sqrt{Z}}\sum_{n}e^{-\beta_{H} E_{n}/2}\varphi_{n}[\phi_{A}]\varphi_{n}[\phi_{B}].
\end{equation}

We are interested in the entanglement entropy of the subsystem $A$. To compute it, one first has to construct the reduced density matrix $\rho_{A}$ and then evaluate the entropy $S_{A} = -\text{Tr}_{A}(\rho_{A}\log{\rho_{A}})$. In practice, however, this can rarely be done analytically. The replica trick allows one to perform the computation as follows:
\begin{equation}
    S_{A} = - \lim_{\alpha \to 1}\partial_{\alpha}\log{\text{Tr}_{A}(\rho_{A}^{\alpha})} = -\lim_{\alpha \to 1}\partial_{\alpha}\log{\left(\frac{Z_{\alpha}}{Z^{\alpha}}\right)},
\end{equation}
where $Z$ is the partition function on $\mathcal{M}$, and $Z_{\alpha}$ is the partition function on the $\alpha$-fold cover $\mathcal{M}_{\alpha}$ of the original manifold, which is constructed by gluing $\alpha$ copies of the space cut along the subsystem $A$ (for integer $\alpha$, followed by an analytic continuation to non-integer values). Importantly, $\mathcal{M}_{\alpha}$ has a conical singularity at the entangling surface $\Sigma = \partial A$ ($\text{codim}(\Sigma) = 2$) with the angle deficit $\delta =2\pi(1-\alpha)$, where $\alpha = \beta/\beta_{H}$, since the Euclidean time now has the period $\tau \to \tau + \beta$, which differs from the original one.

\begin{figure}[H]
    \centering
    \begin{tikzpicture}[scale=2.5]
            \draw[thick] (-0.5,0) -- (0,1) -- (0.5,0);
            
            \draw[thick] (0,-0.15) parabola (0.5,0);
            \draw[thick] (0,-0.15) parabola (-0.5,0);
            \draw[dashed] (0,0.15) parabola (-0.5, 0);
            \draw[dashed] (0,0.15) parabola (0.5, 0);

            \draw[->] (0,-0.3) parabola (0.5,-0.15);
            \draw (0,-0.3) parabola (-0.5,-0.15);
            \node at (0, -0.4) {$\tau$};

            \draw[->] (0.2,1) -- (0.6,0.2);
            \node at (0.7, 0.2) {$r$};
            \node at (0.035,1.1) {$r = r_{h}$};
            \fill (0,1) circle (0.5pt);
            \draw (0,0.65) parabola (0.15,0.7);
            \draw (0,0.65) parabola (-0.15,0.7);
            \node at (0, 0.55) {$\delta$};
    \end{tikzpicture}
    \caption{The manifold $\mathcal{M}_{\alpha}$ with a conical singularity. Near the surface $\Sigma$ it locally looks like $\mathcal{C}_{\alpha}\times\Sigma$, where $\mathcal{C}_{\alpha}$ is a two-dimensional cone.}
    \label{fig:Conical_singularity}
\end{figure}
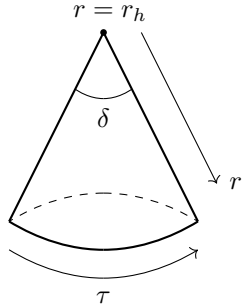

From the technical point of view, the method works as follows. Consider the Euclidean functional integral over metrics $g_{\mu\nu}$ and scalar fields $\phi$ with appropriate boundary conditions, which are implied in what follows and select the required configuration:
\begin{equation}
        Z = \int \mathcal{D}g\mathcal{D}\phi e^{-S[g_{\mu\nu},\phi]}.
 \end{equation}
In the semiclassical approximation, the integral over metrics is dominated by the saddle point $\mathcal{M}$ corresponding to a classical solution of the Einstein equations:
\begin{equation}
     Z \simeq \int \mathcal{D}\phi e^{-(W_{gr}^{bar}[\mathcal{M}]  + W_{matt}[\mathcal{M},\phi])}.
 \end{equation}
In the case of interest, the integral is taken over configurations such that $\mathcal{M}$ has a horizon, and the scalar field is non-minimally coupled to gravity. Then a conical singularity is introduced at the horizon $\Sigma$. As a result, one obtains the manifold $\mathcal{M}_{\alpha}$ with the Euclidean time period $\tau \to \tau + \alpha\beta_{H}$, whose curvature invariants acquire $\delta$-function singularities localized on the horizon \cite{Fursaev:1995ef}:
\begin{equation}
    \begin{gathered}
        R_{\alpha} = R + 4\pi(1-\alpha)\delta(\Sigma),
        \\
        R_{\alpha}^{2} = R^{2} + 8\pi(1-\alpha)\delta(\Sigma)R + \mathcal{O}((1-\alpha)^{2}),
        \\
        (R_{\alpha})_{\mu\nu}(R_{\alpha})^{\mu\nu} = R_{\mu\nu}R^{\mu\nu}  + 4\pi(1-\alpha)\delta(\Sigma)  R_{\mu\nu}n_{i}^{\mu}n_{i}^{\nu} + \mathcal{O}((1-\alpha)^{2}),
        \\
        (R_{\alpha})_{\mu\nu\rho\sigma}(R_{\alpha})^{\mu\nu\rho\sigma} = R_{\mu\nu\rho\sigma}R^{\mu\nu\rho\sigma} + 8\pi(1-\alpha)\delta(\Sigma) R_{\mu\nu\rho\sigma}n^{\mu}_{i}n^{\rho}_{i}n^{\nu}_{j}n^{\sigma}_{j} + \mathcal{O}((1-\alpha)^{2}),        
    \end{gathered}
\end{equation}
where $n_{i}^{\mu}$ ($i=1,2$) are the orthonormal vectors normal to the horizon surface $\Sigma$.

As a result, one obtains the following Euclidean functional integral for the scalar field theory on the background $\mathcal{M}_{\alpha}$:
\begin{equation}\label{GPI}
    \begin{gathered}
        Z_{\alpha} \simeq \int\mathcal{D}\phi e^{-W_{gr}^{bar}[\mathcal{M}_{\alpha}] - W_{matt}[\phi,\mathcal{M}_{\alpha}] }.
    \end{gathered}
\end{equation}
Next, one has to compute the integral over $\phi$ and obtain the total effective action $W_{tot}[\mathcal{M}_{\alpha}] = W_{gr}^{bar}[\mathcal{M}_{\alpha}]  + W_{matt}[\mathcal{M}_{\alpha}]$. Usually one restricts oneself to the one-loop approximation and uses the heat kernel method (which, generally speaking, requires first constructing the heat kernel on the manifold under consideration) to obtain the one-loop effective action \cite{Solodukhin:1995ak}:
\begin{equation}\label{OneLoopMatter}
    \begin{gathered}
        W^{\text{(1-loop)}}_{matt}[\mathcal{M}_{\alpha}] = \frac{1}{2}\log{\text{Det}\left(-\Box_{\mathcal{M}_{\alpha}} + m^{2} + \xi R\right)}  + 2\pi(1-\alpha)\xi\int_{\Sigma}\langle \phi^{2} \rangle + \mathcal{O}((1-\alpha)^{2}),
        \\
       \log{\text{Det}\left(-\Box_{\mathcal{M}_{\alpha}} + m^{2} + \xi R\right)} = -\int_{\epsilon^{2}}^{+\infty}\frac{ds}{s}\text{Tr}\bar{K}_{\mathcal{M}_{\alpha}}(s), \quad \langle \phi^{2}(x) \rangle = \int_{\epsilon^{2}}^{+\infty}ds \bar{K}_{\mathcal{M}_{\alpha}}(s;x,x),
    \end{gathered}
\end{equation}
where $\bar{K}_{\mathcal{M}_{\alpha}}(s;x,x')$ is the heat kernel of the operator $-\Box + m^{2} + \xi R$ on $\mathcal{M}_{\alpha}$ and $\epsilon$ is the UV cutoff. It is important to note that this method allows one to isolate the UV-divergent contributions $W^{div}_{matt}$, which is easily seen from the small-$s$ expansion of the heat kernel, and to perform a consistent renormalization of the UV divergences \cite{Fursaev:1994ea,Solodukhin:1995ak}. For this purpose, the bare gravitational action is chosen in the form
\begin{equation}\label{Grav_matterAction}
    \begin{gathered}
        W_{gr}^{bar}[\mathcal{M}_{\alpha}] = \int_{\mathcal{M}_{\alpha}}\bigg(-\frac{R_{\alpha}-2\Lambda}{16\pi G} + c_{1}R_{\alpha}^{2} + c_{2}(R_{\alpha})_{\mu\nu}(R_{\alpha})^{\mu\nu} + c_{3}(R_{\alpha})_{\mu\nu\rho\sigma}(R_{\alpha})^{\mu\nu\rho\sigma}\bigg),
    \end{gathered}
\end{equation}
where the terms quadratic in curvature are added as counterterms absorbing the UV divergences arising in (\ref{OneLoopMatter}). For example, the leading divergence of the entropy is proportional to the area of $\Sigma$ and can be removed by the standard renormalization of Newton's constant $G_{N}$. Sub-leading divergences must also be renormalized, which leads to the following renormalization condition up to the first order in $(1-\alpha)$:
\begin{equation}\label{renormProcedure}
    \begin{gathered}
         W_{gr}^{ren}(G_{N}^{ren},\Lambda^{ren},c^{ren}_{i}) = W_{gr}^{bar}(G_{N},\Lambda,c_{i}) + W^{div}_{matt}.
    \end{gathered}
\end{equation}
As shown in \cite{Fursaev:1994ea}, this procedure is indeed consistent with the renormalization of the standard UV divergences of the effective action, i.e. at zero order in $(1-\alpha)$. 

After the renormalization we obtain finite effective action, and the entropy takes the form:
\begin{equation}\label{Off_shell_entropy}
    S_{tot}^{con}(\beta_{H}) = \lim_{\alpha \to 1}(\alpha\partial_{\alpha} - 1)W_{tot}[\mathcal{M}_{\alpha}] = S^{con}_{gr}(\beta_{H}) + S^{con}_{matt}(\beta_{H}).
\end{equation}
where:
\begin{equation}
    S_{gr}^{con}(\beta_{H}) = \frac{A}{4G_{ren}} - 8\pi c_{1}^{ren}\int_{\Sigma}R - 4\pi c_{2}^{ren}\int_{\Sigma}R_{\mu\nu}n^{\mu}_{i}n^{\nu}_{i} - 8\pi c_{3}^{ren}\int_{\Sigma}R_{\mu\nu\rho\sigma}n^{\mu}_{i}n^{\rho}_{i}n^{\nu}_{j}n^{\sigma}_{j}.
\end{equation}
In what follows, we will choose a regularization procedure such that $c_i^{ren}=0$.
 
Let us also note the fact that this entropy for the gravitational contribution is calculated without back-reaction; it must either be taken into account separately, or in the replica method one must take into account that at the end of the calculations the temperature should be set equal to the temperature including quantum corrections, as discussed in the Introduction.

\section{On-shell approaches}\label{sec:onshell}
Let us now turn to the computation of the entropy based on the principles of local thermodynamics, see for expmle: \cite{Yokoyama:2023nld}. The main advantage of this approach is that it is formulated entirely on-shell: at every stage the system is in thermal equilibrium at the Hawking temperature, and the entropy is obtained by integrating the first law of thermodynamics along a sequence of equilibrium states. In particular, neither conical singularities nor off-shell geometries are required. Moreover, since the only input is the renormalized stress-energy tensor of the quantum matter, the resulting entropy is automatically finite.

In what follows, we use the following assumptions:
\begin{enumerate}
    \item The system is in local thermodynamic equilibrium.
    
    \item The spacetime background is static and spherically symmetric:
    \begin{equation*}
        ds^{2} = - f(r)dt^{2} + h(r)dr^{2} + r^{2}\gamma_{ij}dx^{i}dx^{j}, \quad i,j = 2,3,\dots,d-1,
    \end{equation*}
    where $\gamma_{ij}$ is the metric on the unit $(d-2)$-sphere.
    \item The Tolman--Ehrenfest relation holds \cite{Tolman:1930ona}:
    \begin{equation*}
        T_{loc}\sqrt{|g_{00}(r,\Omega)|} = T_H =\text{const}.
    \end{equation*}
    \item The expectation value of the stress-energy tensor $\langle T_{\mu\nu}(r,\Omega) \rangle$ is diagonal.
    \item Metric is a function of Hawking temperature $ g_{\mu \nu} (T_H)$.
\end{enumerate}
We admit that some of these assumptions may be rather restrictive, as will be seen below.

The concept of local thermodynamics implies that all extensive quantities (entropy, internal energy, etc.) are replaced by their densities:
\begin{equation*}
    \begin{gathered}
        S \to s = s(r,\Omega, T),\quad U \to u = u(r,\Omega, T),
    \end{gathered}
\end{equation*}
where $s(r,\Omega,T)$ and $u(r,\Omega,T)$ are the entropy and internal energy densities, respectively. At the same time, all intensive quantities (temperature, pressure, etc.) become local:
\begin{equation*}
    \begin{gathered}
        T \to T(r,\Omega),\quad p \to p(r,\Omega).
    \end{gathered}
\end{equation*}

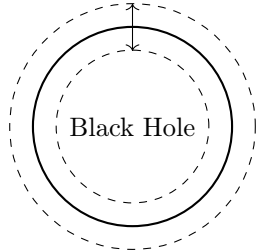
\begin{figure}[H]
    \centering
    \begin{tikzpicture}[scale=2.5]
        \draw[thick] (0,0) circle (15pt);
        \draw[dashed] (0,0) circle (18.5pt);
        \draw[dashed] (0,0) circle (11.5pt);
        \draw[<->] (0,0.40) -- (0,0.65);

        \node at (0,0) {Black Hole};
    \end{tikzpicture}
    \caption{Illustration of the quasi-static process: a slow change of the black hole parameters shifts the horizon and changes the temperature of the system.}
    \label{fig:BHmassVar}
\end{figure}

Consider a quasi-static process in which a parameter of the spacetime, for example the mass of the black hole or the Hubble parameter \cite{Diakonov:2025wtt}, changes slowly, which leads to a change of the temperature of the system. Let us write the first law of thermodynamics in the local form for a volume element with the temperature $T_{loc}$:
\begin{equation}\label{local_first_law}
    T_{loc}\delta s = \delta u + p_r
    \delta\upsilon,
\end{equation}
where $p_{r}$ is the pressure in the radial direction and $\upsilon = \sqrt{h}\,r^{d-2}\sqrt{\det\gamma_{ij}}$ is the proper spatial volume element, so that $\sqrt{|g|} = \upsilon\sqrt{|g_{00}|}$ (here $s$ and $u$ are densities per unit coordinate volume $d^{d-1}x$). We assume that a change of the spacetime parameter deforms the volume element only in the radial direction, so that the work is done only by the radial pressure. Using $u = \rho \upsilon$, where $\rho$ is the energy density, we rewrite the first law and express the variation of the entropy density:
\begin{equation}
    \begin{gathered}
        \frac{1}{\sqrt{|g_{00}|}}T_H\delta s = \delta(\rho\upsilon) + p_{r} \delta\upsilon = \left(\rho + p_{r}\right)\delta\upsilon + \upsilon\delta\rho,
        \\
        T_H \delta s = (\rho + p_{r})\sqrt{|g|}\frac{\delta \upsilon}{\upsilon} + \sqrt{|g|}\delta\rho,
        \\
        \delta s = \frac{\rho + p_{r}}{T_H}\sqrt{|g|}\frac{1}{\upsilon}\frac{\partial \upsilon}{\partial T_H}\delta T_H + \sqrt{|g|}\frac{\partial \rho}{\partial T_H}\frac{\delta T_H}{T_H}.
    \end{gathered}
\end{equation}
In the last line we used that the variations of the spatial volume element and of the energy density are caused only by the change of the spacetime parameter (the mass or the Hubble parameter), and hence of the temperature, since there is a one-to-one correspondence between them. Integrating over the space, we obtain the variation of the entropy:
\begin{equation}
    \begin{gathered}
        \delta S = \int d^{d-1}x \delta s = \frac{\delta T_H}{T_H}\int d^{d-1}x\sqrt{|g|}\frac{\partial \rho}{\partial T_H} + \frac{\delta T_H}{T_H}\int d^{d-1}x\sqrt{|g|}\left(\rho + p_{r}\right)\frac{1}{\upsilon}\frac{\partial\upsilon}{\partial T_H}.
    \end{gathered}
\end{equation}
After integration over the temperature, the entropy takes the form
\begin{equation}
    S^{on-shell}_{matt} = \int^{T_H}\frac{dT_H}{T_H}\int d^{d-1}x\sqrt{|g|}\bigg[\frac{\partial \rho}{\partial T_H} +\left(\rho + p_{r}\right)\frac{1}{\upsilon}\frac{\partial\upsilon}{\partial T_H}\bigg].
\end{equation}
Now, if the spacetime is filled with quantum matter whose renormalized stress-energy tensor $\langle T_{\mu\nu} \rangle$ is known, we can express $\rho$ and $p_{r}$ through its components, $\rho = -\langle T^{0}_{ 0}\rangle$ and $p_{r} = \langle T^{r}_{ r}\rangle$. As a result, the thermodynamic entropy of the quantum matter is given by
\begin{equation}\label{ThermodynEntropyOnShell}
    S^{on-shell}_{matt} = -\int^{T_H}\frac{dT_H}{T_H}\int d^{d-1}x\sqrt{|g|}\bigg[\frac{\partial}{\partial T_H}\langle T^{0}_{ 0}\rangle +\frac{\langle T^{0}_{ 0}\rangle -\langle T^{r}_{ r}\rangle}{\upsilon}\frac{\partial\upsilon}{\partial T_H}\bigg].
\end{equation}
The last expression simplifies considerably when $\langle T_{\mu\nu}\rangle\propto g_{\mu\nu}$ (i.e. $\rho = -p_{r}$): in this case only the first term contributes. This is the case, for example, in de Sitter space \cite{Diakonov:2025wtt}, which we consider in Section~\ref{sec:dS}.   

Next we let us show that \eqref{ThermodynEntropyOnShell} can be written in covariant form \eqref{Our_Formula}. And we start from integration of the ADM mass variation formula:  
\begin{equation}
\label{qwe}
S^{on-shell}_{matt}=\int^{T_H} \frac{d T_H}{T_H} \left(\frac{1}{2} \int \langle T^{\mu\nu} \rangle \frac{\partial g_{\mu\nu} }{\partial T_H}\xi^\rho d V_\rho-\frac{\partial }{\partial T_H} \int \langle T^\mu_\nu \rangle \xi^\nu dV_\mu \right).
\end{equation}
First note that in \cite{Bardeen:1973gs} the deriviation of mass variation formulla impose the condition that when we compare two slightly different solutions we should say which points are chosen to correspond. And make the surfaces of the event horizons the same in the two solutions. That meas that we can change the order of integral and variation in second term of \eqref{qwe}: 
\begin{equation}
\label{qwe1}
S^{on-shell}_{matt}=\int \frac{d T_H}{T_H} \left(\frac{1}{2} \int \langle T^{\mu\nu} \rangle \frac{\partial g_{\mu\nu} }{\partial T_H}\xi^\rho d V_\rho- \int \frac{\partial }{\partial T_H}(\langle T^\mu_\nu \rangle \xi^\nu dV_\mu )\right),
\end{equation}
which actually means that formula \eqref{qwe} must be used carefully and one should write the metric tensor and the radial coordinate in the variable $ \bar{r}=\frac{r}{r_h}$, so that the position of the horizon is always fixed at $\bar{r}=1$. Then one can indeed interchange the integral and the derivative (variation). We note that the matter contribution shifts the horizon, but this contribution is entirely contained in the contribution due to backreaction and the change of the horizon area $\delta \bar{A}$, which, besides the simple shift $\delta A$, also contains the matter contribution. Now, explicitly substituting the expressions for the spherically symmetric metric into \eqref{qwe1} and using covariant conservation laws for SET, one can explicitly obtain that the difference between the covariant expression \eqref{qwe}  and the local one \eqref{ThermodynEntropyOnShell} is a boundary contribution:
\begin{equation}\label{boundaryTerms}
    S^{on-shell}_{matt}\Big|_{\eqref{qwe1}}  -S^{on-shell}_{matt}\Big|_{\eqref{ThermodynEntropyOnShell}}
    =\int\frac{dT_{H}}{T_{H}}\,\frac{\partial \ln r_{h}}{\partial T_{H}}\,\Omega_{d-2}
    \Big[\sqrt{fh}\;r^{\,d-1}(\rho+p_{r})\Big]_{r=r_{h}}^{\,r\to\infty},
\end{equation}
which is equal to zero at infinity and equal to zero at the horizon if $(\rho+p_{r})\Big|_{r=r_{h}}$. This boundary condition is in fact equivalent to demanding that the energy density measured by a freely falling observer remain finite as the observer crosses the horizon \cite{Visser:1992qh}. And this is indeed the cases for de Sitter and BTZ as one can see further.

\section{Example: de Sitter space}\label{sec:dS}
It was shown in \cite{Diakonov:2025wtt} that the off-shell and on-shell methods yield the same entanglement entropy in de Sitter space. This is the simplest example, since the background is maximally symmetric. However, precisely for this reason, the question remains whether this equivalence holds in less symmetric situations and in general. Therefore, in the next section we discuss in detail the computation of the entropy within both approaches for the BTZ black hole. Below we briefly summarize the main steps of the computation of the entanglement entropy in de Sitter space within the two methods; for more details see \cite{Diakonov:2025wtt}. And at the last subsection we will show that the thermodynamic entropy at one loop is determined by the free energy of the full system:
\begin{align}
    S^{TD} =\bar{S}^{BH}+S^{SM}-Q= -\beta_H F^{total} = \log Z^{total}.
\end{align}
For that we will take into account backreaction.  
Note that for de Sitter space we will also use the notation $S^{BH}$ for area law terms, but we mean the area of the cosmological horizon.

\subsection{Off-shell}
The Euclidean functional integral for a scalar field theory in de Sitter space at an arbitrary temperature has been computed in several papers: using the zeta-function regularization \cite{Fursaev:1993hm}, using Harish-Chandra characters \cite{Anninos:2020hfj}, and using the Green function at coincident points \cite{Akhmedov:2021cwh}. The renormalized effective action to first order in $(1-\alpha)$ is given by \cite{Diakonov:2025wtt}:
\begin{equation}\label{renlogZ}
    \log Z^{ren}_{matt} = -W_0\left(T_H\right) - (1-\alpha)W_1\left(T_H\right),
\end{equation}
where $H = 2\pi T_H$ and
\begin{equation}
    \begin{gathered}
        W_0\left(T_H\right) = \frac{17 - 120(2 \nu^{4} + \nu^{2})}{5760} \log{\left(\frac{m^{2}}{H^{2}}\right)} - \frac{1}{6}\zeta'(-1) + \frac {1}{72} - \frac{1}{3} \zeta'(-3) + \frac{1}{288} \nu^{2}\left(6\nu^{2} + 35\right) - 3\xi^{2} -
        \\
        - \frac{1}{8}\left(4\nu^{2} - 7\right) \xi + \frac{\zeta(3)}{8\pi^{2}} - \frac{55}{768} - \textbf{Re}\Bigg[
        \frac{i}{12}\left(4\nu^{3} + \nu\right) \log{\Gamma\left(i\nu + \frac{1}{2}\right)}
        + \frac{1}{24}\left(12\nu^{2} + 1\right) \psi^{(-2)}\left(i\nu + \frac{1}{2}\right) -
        \\
        - \psi^{(-4)}\left(i\nu + \frac{1}{2}\right) + i\nu\psi^{(-3)}\left(i\nu + \frac{1}{2}\right)\Bigg],
    \end{gathered}
\end{equation}
\begin{equation}
    \begin{gathered}
        W_1(T_H) 
        =  \frac{1}{384}\left(4\nu^{2} + 1\right)\left(4 \nu^{2} - 48 \xi + 9\right) \log\left(\frac{m^{2}}{H^2}\right) + \frac{20 \nu^{2} (36 \xi - 7) + 60(31 - 72 \xi) \xi - 199}{1440} - 
        \\
        - \frac{1}{384}\left(4\nu^{2} + 1\right)\left(4 \nu^{2} - 48 \xi + 9\right)\Big[\psi\left(\frac{1}{2} - i \nu\right) + \psi\left(i \nu + \frac{1}{2}\right)\Big],
    \end{gathered}
\end{equation}
where
\begin{equation}
    \nu = \sqrt{\frac{m^{2}}{H^{2}} + 12\xi - \frac{9}{4}}.
\end{equation}

First, one can show that the derivative of the renormalized effective action \eqref{renlogZ} with respect to the inverse temperature yields exactly the energy of the system:
\begin{equation}\label{dS_energy}
    -\Big(\partial_\beta \log Z^{ren}_{matt}\Big) \Bigg|_{\alpha=1} = \langle \hat{E} \rangle,
\end{equation}
where the energy is defined by the stress-energy tensor \cite{Dowker:1975tf, Birrell:1982ix, Bernard:1986vc}, $\langle \hat{E} \rangle = -\int d^{3}x\sqrt{-g}\,\langle T^{0}_{ 0}\rangle$. Hence, in contrast to the standard statistical-mechanical relation, the derivative of the effective action gives $\langle \hat{E} \rangle$ rather than the expectation value of the canonical Hamiltonian $\langle \hat{H} \rangle$. As discussed in the Introduction, these two quantities differ \cite{Iyer:1995kg, Frolov:1998vs}:
\begin{equation}\label{Ehxi}
    \langle \hat{E} \rangle = \langle \hat{H} \rangle - \beta_H ^{-1}\langle \hat{Q} \rangle.
\end{equation}
As a result, the off-shell entropy takes the form
\begin{equation}
    S^{on-shell}_{matt} = -\left(\beta \partial_\beta - 1\right) \log Z^{ren}_{matt} \Big|_{\beta = \beta_{H}} = S^{SM} -\langle \hat{Q} \rangle,
\end{equation}
where $S^{SM} = \beta_{H} \langle \hat{H} \rangle - \beta_{H} F$ is the standard definition of the entropy, which arises from the fact that the Euclidean functional integral $Z^{SM}$ is equal to the partition function $Z^{SM} = \text{Tr}\left[e^{-\beta \hat{H}}\right]$. Explicitly, one finds
\begin{equation}\label{SDS}
    \begin{gathered}
         S^{of-shell} _{matt}= \frac{1}{6}\zeta'(-1)-\frac {1}{72}+\frac{1}{90} \zeta'(-3)-\frac{\nu^{4}}{48}-\frac{7 \nu^{2}}{288} + 6\xi^{2} - \frac{13\xi}{6} - \frac{\zeta(3)}{8\pi^{2}} + \frac{2417}{11520} +
         \\
         + \frac{1}{720} \left(60\nu^{2} (6\xi - 1) + 90\xi - 19\right) \log{\left(\frac{m^{2}}{4\pi^{2} T_{H}^{2}}\right)} + \textbf{Re}\Bigg[ 2\psi^{(-4)}\left(\frac{1}{2} + i\nu\right) - 2i\nu\psi^{(-3)}\left(\frac{1}{2} + i\nu\right) -
         \\
         - \left(\nu^{2} + \frac{1}{12}\right) \psi^{(-2)}\left(\frac{1}{2} + i\nu\right) + \frac{i}{12}\left(4\nu^{3} + \nu\right) \log{\Gamma\left(\frac{1}{2} + i\nu\right)} + \frac{1}{192} \left(4\nu^{2} + 1\right)\left(4\nu^{2} - 48\xi + 9\right) \psi\left(\frac{1}{2} + i\nu\right)\Bigg].
    \end{gathered}
\end{equation}

\subsection{On-shell}
In the off-shell method, in order to obtain the renormalized entropy one has to compute the renormalized effective action to first order in $(1-\beta/\beta_H)$, which is a rather difficult task. The advantage of the on-shell method is that one deals directly with the renormalized stress-energy tensor, which determines the quantum corrections to the entropy. In de Sitter space, the expectation value of the stress-energy tensor of a non-minimally coupled scalar field in the Bunch--Davies state is proportional to the metric tensor \cite{Dowker:1975tf,Birrell:1982ix,Bernard:1986vc}: 
\begin{equation}\label{set}
    \begin{gathered}
        \langle T_{\mu\nu} \rangle = -g_{\mu\nu} \frac{H^{4}}{64\pi^{2}}\Bigg[\left(\frac{1}{4} + \nu^{2}\right)\left(\frac{9}{4} + \nu^{2} - 12\xi\right) \bigg[\psi\left(\frac{3}{2} + i\nu\right) + \psi\left(\frac{3}{2} - i\nu\right) - \log{\left(\frac{m^{2}}{H^{2}}\right)}\bigg] +
        \\
        + \nu^{2} \left(\frac{4}{3} - 12\xi\right) + \xi(72\xi - 19) + \frac{16}{15}\Bigg].
    \end{gathered}
\end{equation}
Since $\langle T_{\mu\nu}\rangle\propto g_{\mu\nu}$, i.e. $\rho_{matt} = -\langle T^{0}_{ 0}\rangle = -p_{r}$, and in the static coordinates $\sqrt{-g}$ does not depend on $T_H$, only the term with the derivative of the energy density survives in \eqref{Our_Formula} (equivalently, in \eqref{ThermodynEntropyOnShell}). As a result, the total quantum-corrected entropy of the Bunch--Davies state takes the form
\begin{equation}\label{dS_onshell}
    S^{on-shell}_{tot} = \frac{A}{4} + \int_{0}^{T_{H}} \frac{dT}{T} \int_{r\leq\frac{1}{2\pi T}} d^{3}x \sqrt{-g}\, \frac{\partial\rho_{matt}(T)}{\partial T},
\end{equation}
where the first term is the standard gravitational contribution, $S^{BH}=\frac{A}{4}$. The second term reproduces exactly the off-shell result \eqref{SDS}.

 In static coordinates $\sqrt{-g}$ does not depend on $T_H$, the first term of \eqref{Our_Formula} vanishes identically, and the entire $T_H$-dependence resides in the integration domain $r\leq(2\pi T_H)^{-1}$. In the rescaled variable $\bar r=r/r_h$, in which the horizon sits at $\bar r=1$ and the derivative may be taken under the integral sign, the domain is fixed and both terms contribute. The two forms agree because their difference is the boundary term \eqref{boundaryTerms}, which vanishes identically in de Sitter: for $\langle T_{\mu\nu}\rangle\propto g_{\mu\nu}$ one has $\rho_{matt}+p_r\equiv 0$.

\subsection{Thermodynamic versus statistical-mechanical entropy}
Let us now check whether the relation between the thermodynamic and statistical-mechanical entropies discussed in the Introduction holds in de Sitter space, i.e. in the presence of a non-vanishing cosmological constant:
\begin{equation}\label{TD_SM_dS}
    S^{TD}=\bar{S}^{BH}+S^{SM}-Q.
\end{equation} 
In this subsection we keep the spacetime dimension $d$ arbitrary (for the case considered above $d=4$). We start from the renormalized effective action \eqref{renlogZ}. The off-shell method then gives
\begin{equation}
    S^{of-shell} _{matt}=-(\alpha \partial_\alpha-1)\log Z^{ren}_{matt} \Big|_{\alpha=1}= -W_0(T_H)-W_1(T_H).
\end{equation}
Using that for de Sitter \cite{Diakonov:2025wtt}: $ S^{of-shell} _{matt}=S^{SM}-Q$, we obtain
\begin{equation}
    S^{SM}-Q=-W_0(T_H)-W_1(T_H),
\end{equation}
where $W_0$ can be expressed as an integral of the matter energy density:
\begin{equation}\label{W0_rho}
    W_0(T_H)=- d \int \frac{d T_H}{T_H}\int_{\mathcal{M}} d^d x \sqrt{g}\, \rho_{matt}(T_H).
\end{equation}
Here $\mathcal{M}\simeq S^{d}$ is the Euclidean de Sitter space of radius $H^{-1}$, which covers the static patch $r\leq (2\pi T_H)^{-1}$ together with the thermal circle of length $\beta_H$. Relation \eqref{W0_rho} follows from the fact that a change of $T_H$ amounts to a constant rescaling of the Euclidean metric, so that $T_H\partial_{T_H}W_0$ is given by the integrated trace of the stress-energy tensor, $\langle T^{\mu}_{ \mu} \rangle = -d\,\rho_{matt}$. On the other hand, within the local thermodynamic approach, $S^{SM}-Q$ can also be expressed through the energy density:
\begin{equation}
    \begin{gathered}
        S^{SM}-Q = \int \frac{d T_H}{T_H}  \int_{r \leq \frac{1}{2 \pi T_H}} d^{d-1} x \sqrt{-g}\, \frac{\partial \rho_{matt}(T_H) }{\partial T_H} =
        \\
        = \int d T_H \frac{\partial}{\partial T_H} \int_{\mathcal{M}} d^d x\sqrt{g}\, \rho_{matt}(T_H)+d \int \frac{d T_H}{T_H}\int_{\mathcal{M}} d^d x \sqrt{g}\, \rho_{matt}(T_H),
    \end{gathered}
\end{equation}
where we used that $\rho_{matt}$ is constant in de Sitter space and that the Euclidean volume scales as $T_H^{-d}$. Comparing the two expressions above, we find
\begin{equation}\label{W1_rho}
    W_1(T_H)=- \int_{\mathcal{M}} d^d x \sqrt{g}\, \rho_{matt}(T_H) = -\beta_H\langle \hat{E} \rangle,
\end{equation}
which is in agreement with \eqref{dS_energy}.

Let us now consider the back-reaction of the quantum matter on the geometry and the resulting correction to the horizon area. From the semiclassical Einstein equations:
\begin{equation}
    R_{\mu\nu} - \frac{1}{2} R g_{\mu \nu} + \Lambda_0 g_{\mu\nu} = 8\pi \langle T_{\mu\nu} \rangle,
\end{equation}
with $\langle T_{\mu\nu} \rangle = -\rho_{matt}\, g_{\mu\nu}$, it follows that the cosmological constant is shifted as
\begin{equation}
    \Lambda=\Lambda_0-\frac{8\pi}{d} \langle T^{\mu}_{ \mu} \rangle = \Lambda_0 + 8\pi\rho_{matt}(T_H).
\end{equation}
As a result, the quantum-corrected Hubble constant is given by
\begin{equation}
\label{habblecorrection}
    H=H_0+\frac{8\pi}{(d-1)(d-2)} \frac{\rho_{matt}(T_H)}{H_0},
\end{equation}
where $\Lambda_0 = \frac{(d-1)(d-2)}{2}H_0^{2}$. Using $A = \Omega_{d-2}H^{-(d-2)}$ and $\text{Vol}(S^{d}) = \frac{2\pi}{d-1}\Omega_{d-2}H^{-d}$, where $\Omega_{n}$ is the area of the unit $n$-sphere, one finds that this shift changes the Bekenstein--Hawking entropy by
\begin{equation}
\label{deltaSeqW1}
    \delta S^{BH}= - \int_{\mathcal{M}} d^d x \sqrt{g}\,  \rho_{matt}(T_H)= W_1(T_H),
\end{equation}
i.e. the correction to the horizon area is expressed through $W_1$.

Hence, for the back-reacted entropy $\bar{S}^{BH} = S^{BH} + \delta S^{BH}$, the right-hand side of \eqref{TD_SM_dS} takes the simple form
\begin{equation}
    \bar{S}^{BH}+S^{SM}-Q = S^{BH} - W_0(T_H).
\end{equation}
On the other hand, in de Sitter space the free energy and the entropy are related as $\beta_H F_{gr}=-S^{BH}$, since the energy of empty de Sitter space vanishes \cite{Galante:2023uyf,Gibbons:1976ue}. The free energy is expressed through the logarithm of the semiclassical partition function, $-\beta_H F^{gr}=-W_{gr}=\log Z_{gr}$, where $W_{gr}$ is the on-shell Euclidean Einstein--Hilbert action, while $-W_0(T_H) = \log Z^{ren}_{matt}\big|_{\alpha=1}$. Therefore, the right-hand side of \eqref{TD_SM_dS} reduces to the logarithm of the full partition function:
\begin{equation}
      S^{TD}=\bar{S}^{BH}+S^{SM}-Q  =\log \left(Z_{gr}Z^{ren}_{matt}\right) =\log Z_{tot}.
\end{equation}
In this sense, the thermodynamic relation in de Sitter space remains the same as at tree level, i.e. the thermodynamic entropy is determined by the free energy of the full system, $S^{TD} = -\beta_H F_{tot} = \log Z_{tot}$, where backreaction is taken into account.

Let us also check that the correction to the area comes from the replica method for the gravitational contribution due to the change of temperature: 
\begin{align}
    \delta S_{BH}=S^{con}_{gr}|_{\beta_H+\delta \beta_H}-S^{con}_{gr}|_{\beta_H}.
\end{align} 
To show that, consider the classical action in the limit \( \alpha \to 1 \), which is given by:
    \begin{gather}
    \label{Wq}
        W_{gr}[\alpha]
        = \\
        =\nonumber
        -\alpha \frac{1}{16 \pi}\int_\mathcal{M} d^{d}x \sqrt{g}  \left(R^{\mathcal{M}}-2 \Lambda\right) - 
        \\ 
        \nonumber
        -
        (1-\alpha) \frac{1}{4}\int_\mathcal{M} d^{d}x \sqrt{g}  \delta\left(\Sigma\right)   +
        O\left((1-\alpha)^2\right),
    \end{gather}
Applying the replica formula, one can find that only the surface term contributes to the entropy: 
\begin{align}S^{con}_{gr}=(\alpha\partial_\alpha-1)W_{gr}[\alpha]=  \frac{A}{4},
    \end{align}
then using that $A = \Omega_{d-2}H^{-(d-2)}$ we can obtain: 
\begin{align}
    \delta S_{BH}=S^{con}_{gr}|_{\beta_H+\delta \beta_H}-S^{con}_{gr}|_{\beta_H} =-(d-2)\frac{\Omega_{d-2}H_0^{-(d-1)}}{4} \delta H
\end{align} 
where $\delta H$ is given by \eqref{habblecorrection}. As a result, one can show that:
\begin{align}
    \delta S_{BH} =-\Omega_{d-2}H_0^{-d} \frac{2\pi}{(d-1)} \rho_{matt}(T_H),
\end{align} 
and using that $\text{Vol}(S^{d}) = \frac{2\pi}{d-1}\Omega_{d-2}H^{-d}_0$, we obtain \eqref{deltaSeqW1}. Thus, indeed, from the replica method for the gravitational contribution, the result is actually the contribution due to back-reaction to the change of the area.

\section{Example: BTZ black hole}
The second example we consider is the BTZ black hole. It differs from the Schwarzschild black hole since it is asymptotically anti-de Sitter rather than asymptotically flat. It has thermodynamic properties like a $(3+1)$-dimensional black hole \cite{Carlip:1995qv}, with Hawking temperature:  
\begin{align}
\label{btztemp}
    T_{H}=\frac{\sqrt{M}}{l} \frac{1}{2\pi} 
\end{align}
Quantum corrections to BTZ entropy were considered in \cite{Ichinose:1994rg,Binosi:1998yu, Mann:1996ze}. Let us note that the answer we obtain later differs from that obtained for example in \cite{Mann:1996ze}, since we take into account the boundary contribution $Q$ due to the conical singularity, whose presence was discussed in detail in Ref. \cite{Solodukhin:1995ak}, and that is needed for a self-consistent renormalization procedure. In addition, this expression is reproduced exactly by the on-shell calculation. We will also not take into account back reaction and derive the total quantum correction, since our main goal of this paper is to check whether the on-shell method indeed reproduces the off-shell result.

In the general case the background we study has the form:
\begin{align}
\label{btzmet}
    ds^2 =-\left(\frac{r^2}{l^2}-M\right) d t^2+\left(\frac{r^2}{l^2}-M\right)^{-1} d r^2 +r^2 d\phi^2.
\end{align}
$l$- the radius of $\mathrm{AdS}_{3}$ which we set to one, $M$ - the mass of the black hole. 

In what follows, we will denote by $B_{3}^{\alpha}$ the Euclidean BTZ with a conical singularity, and by $B_{3}$ the regular Euclidean BTZ without a singularity.

\subsection{Off-shell}

As mentioned earlier, in the off-shell approach one first computes the one-loop effective action (\ref{OneLoopMatter}), knowing the exact expression for the heat kernel, and then finds the entropy (\ref{Off_shell_entropy}). Schematically, this can be represented as: 
\begin{align}
    K^{AdS}\to K^{BTZ}_{\alpha=1} \to K^{BTZ}_{\alpha} \to W_\alpha \to W_\alpha^{ren} \to S_{con}|_{\alpha=1}:
\end{align}
so that for this case one first constructs the heat kernel in anti-de Sitter space, then obtains from it the heat kernel for the BTZ black hole, from that constructs the heat kernel for the space with a conical singularity, computes the effective action, removes all divergences by redefining the constants, and only then computes the entanglement entropy.

For the case of Euclidean BTZ with a conical singularity, the heat kernel was obtained in \cite{Mann:1996ze} and in the non-rotating case for Neumann boundary conditions has the form
\begin{equation}
    \begin{gathered}
        \text{Tr} K_{B_{3}^{\alpha}} = \left( V_{B_{3}^{\alpha}} + 2\pi\alpha A_{+}c_{2}(\alpha) s\right)\frac{e^{-\mu s}}{(4\pi s)^{3/2}} + 2\pi \frac{e^{-\mu s}}{(4\pi s)^{3/2}} A_{+} s\sum_{n=1}^{+\infty}
        \frac{\tanh{\frac{\psi_{n}}{2}}}{\tanh{\frac{\psi_{n}}{2\alpha}}}\frac{e^{-\frac{\psi_{n}^{2}}{4 s}}}{\sinh^{2}{\frac{\psi_{n}}{2}}}
        \\
        c_{2}(\alpha) = \frac{1}{6}\left(\frac{1}{\alpha^{2}}-1\right), \quad \psi_{n} = A_{+}n, \quad  \mu = 1-6\xi.
    \end{gathered}
\end{equation}
where $V_{B_{3}^{\alpha}}$ is the volume of BTZ, $A_{+} = 2\pi r_{+}$ is the area of the horizon surface $\Sigma$. Let us also note that one can consider Dirichlet and transparent boundary conditions. We will not discuss them further.

We restrict ourselves to the non-rotating case, for simplicity of calculation, and also because of the presence of a certain problem with the matter stress-energy tensor for the case of general geometry, which we will mention below. Following the scheme described earlier (\ref{OneLoopMatter}), we need the expression for $\langle \phi^{2} \rangle$, which was obtained in \cite{Lifschytz:1993eb,Shiraishi:1993qnr,Binosi:1998yu} for the case $J = 0$:
\begin{equation}
        \langle \phi^{2} \rangle = \frac{1}{4\sqrt{2}\pi L}\frac{r_{+}}{r}\Bigg(\sum_{n \neq 0}\frac{1}{\sqrt{\cosh{\psi_{n}}-1}} - \sum_{n=-\infty}^{+\infty}\frac{1}{\sqrt{\cosh{\psi_{n}}-1 + 2\left(\frac{r_{+}}{r}\right)^{2}}}\Bigg). 
\end{equation}
After direct computation of the integrals over the heat kernels, we obtain the following result for the one-loop effective action of matter:
    \begin{gather}
        W_{matt}^{con}[B_{3}^{\alpha}] = 
        W_{matt}^{div}[B_{3}^{\alpha}]
   - \\- \nonumber
        \sum_{n=1}^{+\infty}\frac{1}{4n}\frac{\sinh{\left(\frac{A_{+}}{\alpha}n\right)}}{\sinh{ A_{+}n}}\frac{e^{-\sqrt{\mu} A_{+}n}}{\sinh^{2}{\left(\frac{ A_{+}n}{2\alpha}\right)}} + (1-\alpha)\frac{\xi  A_{+}}{\sqrt{2}}\sum_{n=1}^{+\infty}\frac{\sqrt{\cosh{( A_{+}n)}+1} -\sqrt{\cosh{( A_{+}n)}-1}}{\sinh{( A_{+}n)}},
    \end{gather}
where the divergent contribution has the form
\begin{equation}
    W_{matt}^{div}[B_{3}^{\alpha}] = -\frac{1}{(4\pi)^{3/2}}\Bigg[\frac{\alpha}{3\epsilon^{3}}\int_{B_{3}}1 + \frac{\alpha}{\epsilon}\left(\frac{1}{6}-\xi\right)\int_{B_{3}}R + \frac{4\pi}{\epsilon}(1-\alpha)\left(\frac{1}{6}-\xi\right)\int_{\Sigma}1\Bigg].
\end{equation}

Let us carry out the renormalization procedure described in detail in \ref{sec:offshell} of this work, by redefining the bare constants
\begin{equation}
    \begin{gathered}
        W_{tot} = W_{gr}^{bar}(G_{3},\Lambda) + W^{div}_{matt} + W^{fin}_{matt} = W_{gr}^{ren}(G_{3}^{ren},\Lambda^{ren}) + W^{fin}_{matt},
        \\
        \frac{1}{16\pi G_{3}^{ren}} = \frac{1}{16\pi G_{3}} + \frac{\frac{1}{6}-\xi}{(4\pi)^{3/2}}\frac{1}{\epsilon}, \quad \frac{\Lambda^{ren}}{8\pi G_{3}^{ren}} = \frac{\Lambda}{8\pi G_{3}} -\frac{1}{3(4\pi)^{3/2}}\frac{1}{\epsilon^{3}}.
    \end{gathered}
\end{equation}
Hence the renormalization procedure is self-consistent in zero and first order in $(1-\alpha)$.
As a result, we obtain the following renormalized one-loop effective action
    \begin{gather}
        W_{tot}[B_{3}^{\alpha}] = W_{gr}^{ren}[B_{3}^{\alpha}]
        -  \\- \nonumber
        \sum_{n=1}^{+\infty}\frac{1}{4n}\frac{\sinh{\left(\frac{ A_{+}}{\alpha}n\right)}}{\sinh{ A_{+}n}}\frac{e^{-\sqrt{\mu} A_{+}n}}{\sinh^{2}{\left(\frac{ A_{+}n}{2\alpha}\right)}} + (1-\alpha)\frac{\xi  A_{+}}{\sqrt{2}}\sum_{n=1}^{+\infty}\frac{\sqrt{\cosh{( A_{+}n)}+1} -\sqrt{\cosh{( A_{+}n)}-1}}{\sinh{( A_{+}n)}},
        \\ \nonumber
        W_{gr}^{ren}[B_{3}^{\alpha}] = -\frac{\alpha}{16\pi G_{3}^{ren}}\int_{B_{3}}\left(R - 2\Lambda^{ren}\right) - (1-\alpha)\frac{A_{+}}{4G_{3}^{ren}},
    \end{gather}
knowing which, we can compute the desired entropy by the formula (\ref{Off_shell_entropy})
\begin{equation}\label{BTZ_off_shell_answer}
    \begin{gathered}
        S^{con}_{gr}+S^{con}_{matt} = \frac{A_{+}}{4G_{3}^{ren}} +\sum_{n=1}^{+\infty}s_{n} -\xi  A_{+}\sum_{n=1}^{+\infty}\frac{e^{-\frac{1}{2} A_{+}n}}{\sinh{( A_{+}n)}}
        \\
        s_{n} = \frac{1}{2n}\frac{e^{-\sqrt{\mu} A_{+}n}}{\cosh{( A_{+}n)} -1}\left(1 +  A_{+}n\coth{( A_{+}n)} - \frac{ A_{+}n\sinh{( A_{+}n)}}{\cosh{( A_{+}n})-1}\right)
    \end{gathered}
\end{equation}
where $\xi = \frac{1}{8}$, $\mu = \frac{1}{4}$.  This term again does not take into account back reaction since in in one loop approximation one should take into account that we have a quantum correction to the Bekenstein-Hawking temperature and it modify only the gravitational part $S^{con}_{gr}$, since the matter contribution $S^{con}_{matt}$ is already the leading correction:

\subsection{On-shell}
Without further ado, let us turn directly to the computation of the on-shell entropy. For the non-rotating BTZ case, the renormalized expression for the stress-energy tensor is known from Ref. \cite{Lifschytz:1993eb,Shiraishi:1993qnr,Binosi:1998yu}.

    \begin{gather}
        \langle T_{ \nu}^{\mu}\rangle = \\=\nonumber \frac{1}{16\pi L^{3}r^{3}}\sum_{n=1}^{+\infty}\Bigg[\left(\frac{r_{+}^{2}}{2f_{n}}\bigg(1 - \frac{1}{\left(1 + (f_{n}r)^{-2}\right)^{3/2}}\bigg) +\frac{1}{f_{n}^{3}}\right)\text{diag}(1,1,-2) 
        - \frac{3}{2f_{n}^{3}}\frac{1-\frac{r_{+}^{2}}{r^{2}}}{\left(1 + (f_{n}r)^{-2}\right)^{5/2}}\text{diag}(1,0,-1)\Bigg],
        \\ \nonumber
        f_{n} = \frac{1}{r_{+}}\sinh{\left(\frac{ A_{+}n}{2}\right)}, \quad \quad r_{+} = 2\pi T_{H}.
    \end{gather}
    
After substituting the explicit form of the geometry, expression \eqref{Our_Formula} takes the following form 
\begin{equation}
    \begin{gathered}
        S_{matt}^{on-shell} = \int \frac{dT_{H}}{T_{H}}\bigg(\frac{1}{2}\int d^{d-1}x\sqrt{-g}\langle T^{\mu}_{ \rho}\rangle g^{\rho\nu}\frac{\partial g_{\mu\nu}}{\partial T_{H}} - \frac{\partial}{\partial T_{H}}\int d^{d-1}x\sqrt{-g}\langle T^{0}_{ 0}\rangle\bigg) =
        \\
        =  2\pi \int \frac{dT_{H}}{T_{H}} \int_{r_{+}}^{+\infty}dr \ r\Bigg(\frac{r_{+}}{r^{2}-r_{+}^{2}}\left(\langle T^{1}_{ 1}\rangle - \langle T^{0}_{ 0}\rangle \right)\frac{\partial r_{+}}{\partial T_{H}} - \frac{\partial}{\partial T_{H}}\langle T^{0}_{ 0}\rangle \Bigg).
    \end{gathered}
\end{equation}
where we use that the metric \eqref{btzmet} is a function of temperature \eqref{btztemp}: 
\begin{align}
    g_{\mu \nu} (M )=
    g_{\mu \nu} \left((2\pi T_H)^2 \right) ,
\end{align}
and that $r_+ =2\pi T_H$.
Omitting the calculations we arrive at the following answer for the total BTZ entropy 
\begin{equation}
    S_{matt}^{on-shell} =\sum_{n=1}^{+\infty}s_{n} -\xi  A_{+}\sum_{n=1}^{+\infty}\frac{e^{-\frac{1}{2} A_{+}n}}{\sinh{( A_{+}n)}}
\end{equation}
which in agreement with matter part of (\ref{BTZ_off_shell_answer}).

Both conditions required for the equivalence of \eqref{qwe1} and \eqref{ThermodynEntropyOnShell}, i.e. the vanishing of the boundary term \eqref{boundaryTerms}, are satisfied here.

Thus, using the example of BTZ space, which is less symmetric than de Sitter space, we have shown that the generalized entropy can be computed using two different approaches.

\section{Discussion}

In this paper, we have studied the entanglement entropy of black holes from the perspective of the first law of thermodynamics. We have developed an on-shell method that allows one to compute the entanglement entropy of quantum matter directly from the renormalized stress-energy tensor, without of the off-shell techniques such as the replica trick. This approach is valid for stationary axisymmetric black holes and de Sitter space elaborate a finite, renormalized result.

We have shown that the on-shell entropy coincides with the off-shell conical entropy, including the contact term $Q$ that arises for non-minimally coupled fields. This equivalence was illustrated with two explicit examples: de Sitter space and the BTZ black hole. In de Sitter space, we also demonstrated that the thermodynamic entropy at one loop is determined by the free energy of the full system, $S^{TD} = \log Z_{tot}$, once backreaction is taken into account. For the BTZ black hole, we computed the entropy both off-shell (using the heat kernel on a cone) and on-shell (using the renormalized stress-energy tensor), and found perfect agreement.

Our results suggest that black hole entanglement entropy can be understood as a thermodynamic quantity, extensive in the volume of space, rather than merely a boundary term. This provides a new perspective on the nature of black hole entropy and its relation to the statistical mechanics of the underlying microstates.

Several open questions remain. The most important is a general proof of the equivalence between the on-shell and off-shell methods for arbitrary geometries and field theories. While we have checked the equivalence in two examples, a general proof would be desirable. It would also be interesting to extend our analysis to charged and rotating black holes, as well as to higher-derivative gravity theories. Another important direction is to understand the role of self-interactions and higher-order corrections, where the renormalization procedure may become more subtle. We leave these questions for future work.

\section*{Acknowledgments}
The authors would like to thank Dmitry S. Ageev, and Emil T. Akhmedov for valuable discussion. GSZ would like to thank Zhenbin Yang for a fruitful conversation about topics of the article. The work of Dmitrii Diakonov was supported by the grant from the Foundation for the Advancement of Theoretical Physics and Mathematics “BASIS”, and by the state assignment of the Institute for Information Transmission Problems of the RAS.

\newpage

\bibliography{bib.bib}
\bibliographystyle{utphys.bst}
\end{document}